\documentclass[journal=jacsat,manuscript=article]{achemso}
\usepackage{chemformula}
\usepackage[T1]{fontenc}

\usepackage{amsmath,soul}
\usepackage{graphicx}

\title{Noncollinear electric dipoles in a polar, chiral phase of CsSnBr$_3$ perovskite}
\author{Douglas H. Fabini}
\affiliation{Max Planck Institute for Solid State Research, 70569 Stuttgart, Germany}
\alsoaffiliation{Department of Chemistry, Massachusetts Institute of Technology, Cambridge, Massachusetts 02139, United States}
\email{fabini@mit.edu}
\author{Kedar Honasoge}
\affiliation{Max Planck Institute for Solid State Research, 70569 Stuttgart, Germany}
\author{Adi Cohen}
\affiliation{Department of Chemical and Biological Physics, Weizmann Institute of Science, Rehovot 76100, Israel}
\author{Sebastian Bette}
\affiliation{Max Planck Institute for Solid State Research, 70569 Stuttgart, Germany}
\author{Kyle M. McCall}
\affiliation{Department of Chemistry, Northwestern University, Evanston, Illinois 60208, United States}
\alsoaffiliation{Present Address: Department of Materials Science and Engineering, University of Texas at Dallas, Richardson, Texas 75080, United States}
\author{Constantinos C. Stoumpos}
\affiliation{Department of Materials Science and Technology, University of Crete, Vassilika Voutes, 70013 Heraklion, Greece}
\author{Steffen Klenner}
\affiliation{Institut f{\"u}r Anorganische und Analytische Chemie, Universit{\"a}t M{\"u}nster, 48149 M{\"u}nster, Germany}
\author{Mirjam Zipkat}
\affiliation{Department of Chemistry, Ludwig-Maximilians-Universit{\"a}t, 81377 M{\"u}nchen,Germany}
\author{Le Phuong Hoang}
\affiliation{Max Planck Institute for Solid State Research, 70569 Stuttgart, Germany}
\author{J{\"u}rgen Nuss}
\affiliation{Max Planck Institute for Solid State Research, 70569 Stuttgart, Germany}
\author{Reinhard K. Kremer}
\affiliation{Max Planck Institute for Solid State Research, 70569 Stuttgart, Germany}
\author{Mercouri G. Kanatzidis}
\affiliation{Department of Chemistry, Northwestern University, Evanston, Illinois 60208, United States}
\author{Omer Yaffe}
\affiliation{Department of Chemical and Biological Physics, Weizmann Institute of Science, Rehovot 76100, Israel}
\author{Stefan Kaiser}
\affiliation{Max Planck Institute for Solid State Research, 70569 Stuttgart, Germany}
\author{Bettina V. Lotsch}
\affiliation{Max Planck Institute for Solid State Research, 70569 Stuttgart, Germany}
\alsoaffiliation{Department of Chemistry, Ludwig-Maximilians-Universit{\"a}t, 81377 M{\"u}nchen,Germany}

\begin{tocentry}
  \centering
  \includegraphics[width=3.25in]{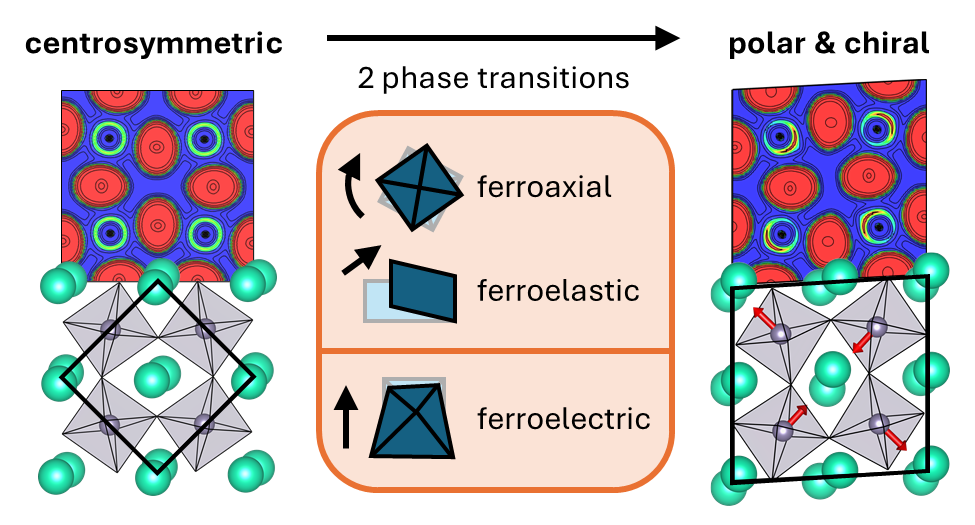}
\end{tocentry}

\begin{document}

\maketitle

\newpage

\begin{abstract}
Polar and chiral crystal symmetries confer a variety of potentially useful functionalities upon solids by coupling otherwise noninteracting mechanical, electronic, optical, and magnetic degrees of freedom.
We describe two unstudied phases of the 3D perovskite, CsSnBr$_3$, which emerge below 85~K due to the formation of Sn(II) lone pairs and their interaction with extant octahedral tilts.
Phase II (77~K<$T$<85~K, space group $P2_1/m$) exhibits ferroaxial order driven by a noncollinear pattern of lone pair-driven distortions within the plane normal to the unique octahedral tilt axis, preserving the inversion symmetry observed at higher temperatures.
Phase I ($T$<77~K, space group $P2_1$) additionally exhibits ferroelectric order due to distortions along the unique tilt axis, breaking both inversion and mirror symmetries.
This polar and chiral phase exhibits second harmonic generation from the bulk and a large, intrinsic polarization$-$electrostriction coefficient along the polar axis ($Q_{22}\approx$~1.1~m$^4$~C$^{-2}$), resulting in acute negative thermal expansion ($\alpha_V=-9\times10^{-5}$~K$^{-1}$) through the onset of spontaneous polarization.
The unprecedented structures of phases I and II were predicted by recursively following harmonic phonon instabilities to generate a tree of candidate structures (``first-principles phonon mode mapping'') and subsequently corroborated by synchrotron X-ray powder diffraction and polarized Raman and $^{81}$Br nuclear quadrupole resonance spectroscopies.
Phase I appears to exhibit the most complex structure reported for a commensurately modulated (single) perovskite, as measured by the index of the space group in that of the cubic aristotype (192).
Relativistic electronic structure scenarios compatible with reported photoluminescence measurements are discussed, and preliminary attempts to suppress unintentional hole doping to allow for ferroelectric switching are described. Together, the polar symmetry, small bandgap, large spin-orbit splitting of Sn 5$p$ orbitals, and predicted strain sensitivity of the symmetry-breaking distortions suggest bulk samples and epitaxial films of CsSnBr$_3$ or its neighboring solid solutions as strong candidates for bulk Rashba effects.

\end{abstract}

\newpage

\section{Introduction}

The impacts of spin-orbit coupling (SOC) on the electronic structure of crystals have been studied and exploited since Elliott, Dresselhaus, Rashba, and colleagues observed that otherwise spin-degenerate bands split under the influence of certain broken spatial symmetries.\cite{Bihlmayer2015} In noncentrosymmetric crystals and those which are additionally polar, the spin-orbit interaction leads to a range of useful couplings between charge, spin, and polarized light which can be harnessed in information processing and sensing devices.\cite{Datta1990, Murakami2003, Sinova2004}

While large relativistic spin-splittings have been observed on metal surfaces\cite{Ast2007} and in engineered heterostructures,\cite{Das1990} relatively few semiconductors have been investigated in detail which exhibit bulk Rashba effects, though a recent screening effort identifies many candidates among known inorganic crystals.\cite{MeraAcosta2020} The two most studied such systems are mixed-anion BiTeI,\cite{Ishizaka2011, Murakawa2013} in which polarity derives from alternating layers of the distinct anions, rendering the polarization and spin texture fixed, and GeTe,\cite{DiSante2012, Liebmann2015, Li2021, Kremer2020} in which ferroelectric switching is possible in principle but challenging in practice.\cite{Rinaldi2018} Both of these systems exhibit narrow (<1~eV) bandgaps and associated challenges with control of electronic doping.

New bulk Rashba semiconductors have recently been sought among perovskite halides of the heavy main group metals. The importance of the spin-orbit interaction in the electronic structure and optical properties of perovskite halides was recognized and examined theoretically by Even and coworkers.\cite{Even2013} Shortly thereafter, others proposed that Rashba or Dresselhaus effects occur at the band edges of hybrid organic$-$inorganic lead(II) perovskites due to parallel alignment of dipolar molecular cations breaking inversion symmetry.\cite{Kim2014, Stroppa2014, Brivio2014} A number of theories followed regarding the impacts that large bulk Rashba effects would have on optical absorption and radiative recombination processes.\cite{Zheng2015, Kepenekian2015, Azarhoosh2016, Etienne2016} Niesner and coworkers reported a large Rashba effect in the valence band of CH$_3$NH$_3$PbBr$_3$ from angle-resolved photoemission spectroscopy (ARPES)\cite{Niesner2016}, whereas others reported that such an effect does not exist in this or related compounds.\cite{Frohna2018, Che2018, Sajedi2020} Far fewer studies of relativistic effects on perovskite halides focus on inorganic compounds where the complexity associated with low symmetry molecular cations is not present and there is little ambiguity about crystallographic symmetry\cite{Frohna2018} or representing the material in ground state atomistic models.\cite{Sajedi2020, MeraAcosta2020, Bhumla2021} Between the pure Rashba or Dresselhaus limits, an appropriate balance between SOC parameters can lead to technologically useful persistent spin textures,\cite{Schliemann2003} as computed for a layered Pb(II) chloride.\cite{Jia2020}

The electron configuration of Pb(II), Sn(II), or Bi(III) in perovskite halides of the heavy main group is a defining feature of these materials, impacting significantly their electronic structures and dielectric properties,\cite{Fabini2020} and offering the opportunity to selectively break symmetries via formation of a lone pair. First-principles studies leading to the modern theory of lone pair stereochemical activity\cite{Watson1999, Waghmare2003, Walsh2005, Stoltzfus2007} reveal that a lone pair is formed on an $ns^2$ cation when the energetic benefit of mixing between anion $p$ orbitals and cation $s$ and $p$ orbitals (which interact under acentric distortions of the cation environment) exceeds the energetic penalty of reduced coordination.\cite{Walsh2011} This distortion from high symmetry, an example of the general pseudo-Jahn$-$Teller effect,\cite{Bersuker2013} is thus strongly dependent on the composition of both cation and anion,\cite{Walsh2011} and can be indirectly tuned in perovskites via the size of the A-cation.\cite{Fabini2016, Laurita2017}

Lone pair formation on the octahedral site is the mechanism of inversion symmetry-breaking in the small number of 3D perovskite main group halides which are unambiguously polar and which neatly illustrate the composition-dependence of lone pair formation above: CsGe$X_3$ ($X$ = Cl, Br, I)\cite{Thiele1987}, RbGeBr$_3$,\cite{Thiele1988} RbGeI$_3$ at elevated temperature,\cite{Thiele1989} CH(NH$_2$)$_2$Ge$_{0.5}$Sn$_{0.5}$Br$_3$,\cite{Liu2022} CsPbF$_3$ at low temperature,\cite{Berastegui2001} and CH$_3$NH$_3$SnBr$_3$ at low temperature.\cite{Swainson2010} In contrast, excepting (methylhydrazinium)PbBr$_3$,\cite{Maczka2020} perovskite Pb(II) iodides and bromides are centrosymmetric, with no stereochemically expressed lone pair. Moving to perovskite-related layered systems relaxes some of the size and shape constraints of 3D systems, and inversion symmetry can be broken through the incorporation of low-symmetry spacer cations, leading to a bulk Rashba effect.\cite{Zhai2017, Schmitt2020}

Much as broken inversion symmetry can confer enhanced functionality via relativistic effects, broken mirror symmetry leads to other useful couplings between charge, spin, and polarized light.\cite{Ray1999, Naaman2012} Chiral responses have been reported for Pb(II) halide nanocrystals capped with chiral ligands\cite{He2017} and layered perovskite-related materials with chiral spacer cations.\cite{Long2018, Lu2019} Very few chiral 3D perovskites of the main group have been reported: (guanidinium)$_{0.5}$(1,2,4-triazolium)$_{0.5}$SnI$_3$, a red and somewhat unstable chiral Sn(II) iodide, crystallizes with $P4_32_12$ (\#96) space group symmetry.\cite{McNulty2020}

Based on the theory of lone pair stereochemical activity and the accumulated observations on neighboring compounds, we sought to identify inorganic perovskites with overlooked polar ground state structures, particularly those heavier than the polar Ge$^{2+}$ halides\cite{Thiele1987} and with narrower bandgaps than CsPbF$_3$.\cite{Berastegui2001} If polar, black CsSnI$_3$ would be expected to exhibit a stronger bulk Rashba effect in the valence band than a lighter bromide, but it is known to interconvert rapidly to a yellow, 1D phase\cite{Yamada1991, Chung2012} and the only published low temperature measurements at the time revealed monotonic evolution of the photoluminescence\cite{Yu2011} which does not suggest unreported phase transitions. Red CH$_3$NH$_3$SnBr$_3$ adopts a polar, monoclinic structure below 230~K but has the added complexity of a low-symmetry molecular cation and an unsolved, likely triclinic, phase below 188~K.\cite{Swainson2010} Given the indirect influence of A-cation size\cite{Laurita2017}, we hypothesized that black CsSnBr$_3$ may exhibit a polar ground state at yet lower temperatures. In fact, two reports hinted at the possibility of an unstudied phase transition in CsSnBr$_3$ at $\sim$85~K by $^{81}$Br nuclear quadrupole resonance (NQR) spectroscopy\cite{Yamada1988} and, concurrent with this work, Raman spectroscopy from some of the present authors.\cite{Gao2021} While NQR indicated a second-order transition at $\sim$85~K to a phase with at least seven distinct Br sites, no further details about the structure, symmetry, or properties of this phase were reported.\cite{Yamada1988} Tantalizingly, this compound exhibits the narrowest bandgap of (single) perovskite group 14 bromides (an important predictor of Rashba effect strength),\cite{Bahramy2011} and its inorganic composition has allowed for the growth of epitaxial thin films.\cite{Wang2017, Wang2017-2, Zhang2021}

CsSnBr$_3$ has repeatedly attracted attention since reports of its first preparations, observation of substantial electronic conductivity, and measurement of M{\"o}ssbauer spectra.\cite{Donaldson1973, Donaldson1973-2, Scaife1974} Several studies sought to understand its electronic structure in light of the combination of bright luminescence, electronic conductivity, and gapless band structure in low level theoretical treatments, with modern first-principles approaches providing clarification.\cite{Clark1981, Bose1993, Huang2013} Recent studies have examined CsSnBr$_3$ in semiconducting applications\cite{Gupta2016, Li2018, Li2019, Zhang2020, Xie2020} or clarified its structure evolution and the property impacts of proximity to lone pair formation.\cite{Fabini2016} Most recently, photoluminescence measurements revealed multipeaked emission and a large blue-shift on cooling below 70~K,\cite{Tan2023} and screened hybrid functional calculations aimed at improving the description of lone pairs were used to predict an unrealized polar, monoclinic phase for CsSnBr$_3$ and CsSnI$_3$ with $Pc$ symmetry.\cite{Swift2023}

Here, we identify and describe the two low temperature phases of CsSnBr$_3$ which emerge due to formation of Sn(II) lone pairs. Phase II (77~K<$T$<85~K, space group $P2_1/m$, \#11) exhibits ferroaxial order and a new electric polarization texture for perovskites, characterized by a noncollinear arrangement of acentric Sn(II) environments with dipoles pointing in the plane normal to the unique octahedral tilt axis. In phase I ($T$<77~K, space group $P2_1$, \#4), the lone pairs additionally drive ferroic (parallel) distortions along the unique tilt axis, breaking inversion and mirror symmetries and resulting in a polar, chiral phase which exhibits second harmonic generation from the bulk and acute negative thermal expansion along the polar axis via electrostrictive coupling to the spontaneous polarization. The new structure types observed for these phases were identified by first-principles phonon mode mapping and corroborated by synchrotron powder diffraction and polarized Raman and $^{81}$Br NQR spectroscopies. Phase I appears to exhibit the most complex structure reported for a commensurately modulated (single) perovskite, as measured by the number of symmetry elements and density of lattice points with respect to those of the aristotype. Elevated electronic conductivity in current samples ($\sim10^2$~$\Omega$~cm at room temperature) precludes ferroelectric switching of the polarization, and experimental and theoretical comparison to CsGeBr$_3$ and CsPbBr$_3$ homologues reveals the conductivities of the compounds closely track the ionization potentials, likely via their influence on the formation energy of $p$-type defects. Hybrid functional electronic structure calculations indicate a substantial spin-orbit splitting of Sn 5$p_{1/2}$ and 5$p_{3/2}$ states at the conduction band minimum ($\sim$0.5~eV), suggesting enhanced polarization via tensile strain in epitaxial thin films as a viable route to enhanced Rashba effects and associated transport phenomena in this inorganic, 3D perovskite.

\section{Results \& Discussion}

\begin{figure}
  \centering
    \includegraphics[width=0.5\textwidth]{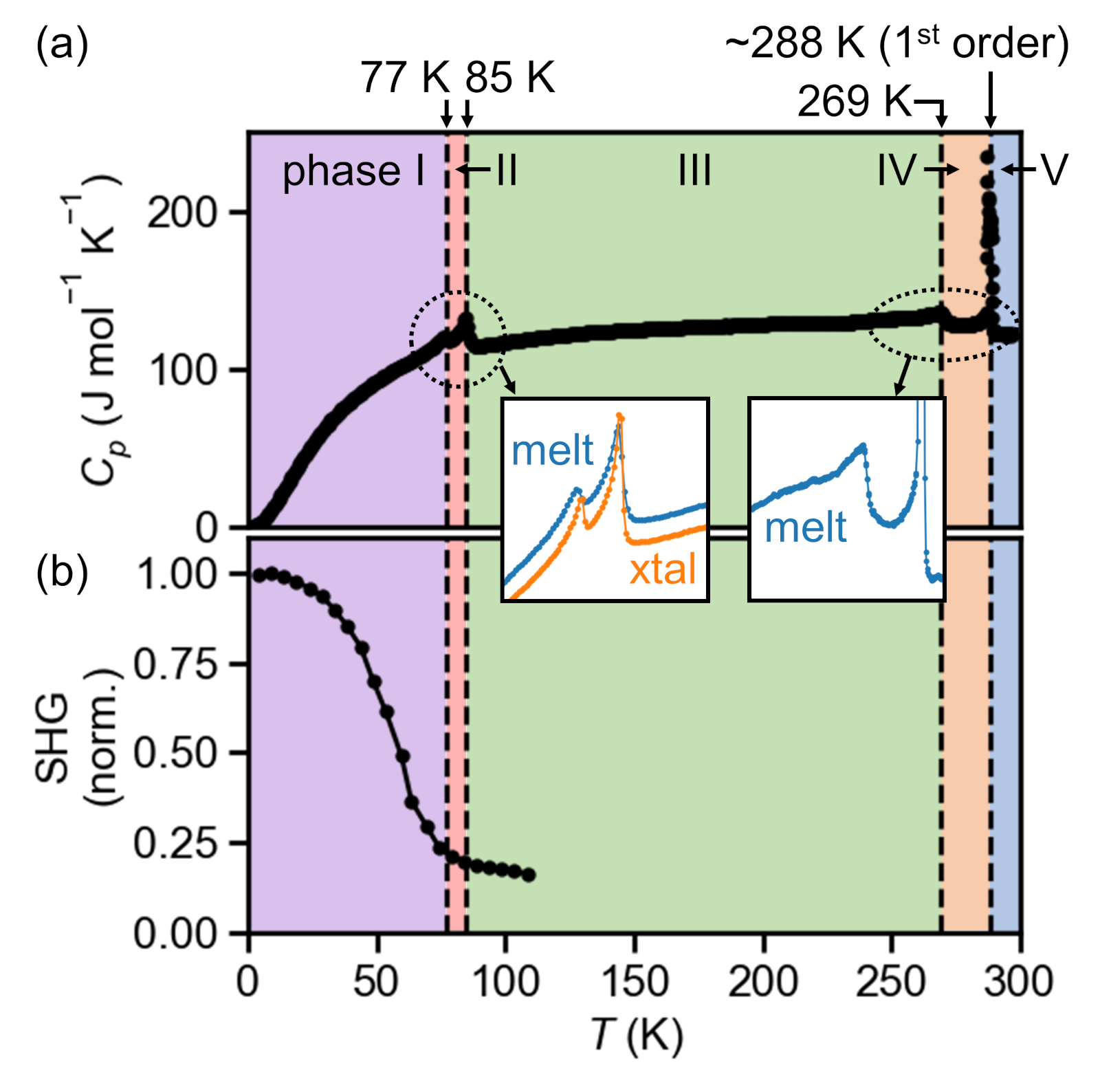}
  \caption{Existence of a noncentrosymmetric ground state phase of CsSnBr$_3$.
  (a) Specific heat, $C_p$, revealing phase transitions between five solid phases, numbered I$-$V starting from low temperature and indicated by colored shading. Transition temperatures ($\pm$1~K) are annotated. The inset shows the transitions to phases I and II are robust to preparation route (``melt'': polycrystalline ingot from the melt; ``xtal'': single crystal grown from solution).
  (b) Intensity of second harmonic generation (SHG) from a single crystal, indicating loss of inversion symmetry in phase I. ``norm.'' = normalized to maximum value.}
  \label{Fig:Cp-SHG}
\end{figure}

\textbf{Existence of a noncentrosymmetric ground state phase of CsSnBr$_3$.} Heat capacity measurements of CsSnBr$_3$ reveal four solid$-$solid phase transitions (Figure~\ref{Fig:Cp-SHG}a). We number the phases I$-$V starting from low temperature. Apart from the first-order transition between phases IV and V, all other transitions are continuous (in this case, second-order), placing restrictions on the symmetry relationships between phases I$-$IV. Phases III$-$V have been the subject of much study, including of their crystal structures.\cite{Mori1986,Fabini2016} While the existence of phase II was evident from $^{81}$Br nuclear quadrupole resonance \cite{Yamada1988} and Raman spectroscopies,\cite{Gao2021} no other information about this phase, or about the existence of an additional, lower temperature phase (I), has been reported. We note that the existence of these unstudied phases is robust to preparation route, with polycrystalline ingots solidified from the melt or single crystals grown from organic solvent (Figure~S1) displaying matching phase transition temperatures within $\pm$1~K (Figure~\ref{Fig:Cp-SHG}a). While the intermediate temperature regime (200~K<$T$<300~K) is not the focus of this work, we note that these data (as well as group theoretic analysis, \textit{vide infra}) are incompatible with the chiral, second tetragonal phase claimed by Mori and Saito (Figure~S2).\cite{Mori1986}

Measurements of second-harmonic generation (SHG) from single crystals of CsSnBr$_3$ reveal a pronounced enhancement of the nonlinear optical susceptibility in phase I, implying a loss of structural inversion symmetry (Figure~\ref{Fig:Cp-SHG}b). The background level observed at higher temperatures likely derives from surfaces, extended defects, and possibly higher order electric quadrupole transitions. The presence of phases I and II is additionally supported by $^{119}$Sn M{\"o}ssbauer spectroscopy which shows broadening of the signal on cooling, indicating unresolved site splitting, unresolved quadrupolar splitting, or both (Figure~S3 and Table~S1).

\textbf{Solving the crystal structures of CsSnBr$_3$ phases I and II.} Attempts to solve the unknown structures of phases I and II with laboratory X-ray diffraction (XRD) were unsuccessful. Laboratory Cu-K$\alpha$ powder XRD gave inadequate resolution to resolve subtle Bragg peak splittings and inadequate signal to detect weak superstructure reflections. Laboratory single crystal XRD measurements with Mo-K$\alpha$ radiation could detect the different phase transitions at the appropriate temperatures. Nevertheless, as the consecutive phase transitions were accompanied by multiple twinning, the crystal of phase III ($Pnma$) had to be handled as a twin with three meaningful volume fractions, already. Phase I and II could be understood in terms of ``twins of twins,'' which lead to peak splitting because of the monoclinic system, and their structures could not be solved due to the resolution limit of the laboratory equipment. Nonetheless, we confirmed the structures of phases III$-$V\cite{Fabini2016} with single crystal XRD, and the structure models have been deposited in the Inorganic Crystal Structure Database (FIZ Karlsruhe, 76344 Eggenstein-Leopoldshafen, Germany, deposition numbers 2302805, 2302804, and 2302966). Faced with limited synchrotron access worldwide in 2020, we endeavored to use theoretical methods and symmetry relationships to determine the unknown structures.

We constructed a tree of candidate structures and phase transitions starting from the experimentally known structure of phase III by recursively computing the harmonic phonon dispersion of each phase (with interatomic force constants from density functional theory, DFT) and following unstable phonon eigenvectors or linear combinations thereof to new saddle points or local minima on the potential energy surface. This approach, recently termed ``first-principles phonon mode mapping,''\cite{Rahim2020} has been employed to predict individual instabilities\cite{Lin2019} or to retroactively understand or clarify the experimentally observed phase evolution of complex materials like the perovskite, CsPbF$_3$,\cite{Berastegui2001, Smith2015} and the pyrochlore, Bi$_2$Sn$_2$O$_7$\cite{Lewis2016, Rahim2020}. This study appears to be among the first to use first-principles phonon mode mapping to successfully predict the crystal structure evolution of a material across multiple phase transitions before it is experimentally known (\textit{vide infra}).  We find a new type of structural distortion for perovskites, resulting in two new structure types which combine octahedral tilting in three dimensions with lone pair-driven distortions which are noncollinear in two dimensions and ferroic (parallel) in the third.

The harmonic phonon band structure of CsSnBr$_3$ phase III computed by DFT is sensitive to the exchange and correlation functional and the inclusion of dispersion corrections (Figure~S4). The approach which correctly finds unstable modes, the generalized gradient approximation functional of Perdew, Burke, and Ernzerhof (GGA-PBE)\cite{Perdew1996} without dispersion corrections, finds several unstable phonon modes at many locations throughout the Brillouin zone, including sizable volumes enclosing some high symmetry points or the lines or planes joining them. Due to the large number of unstable modes and (technically infinite) number of distinct wavevectors, we made simplifying assumptions: Only instabilities at high symmetry wavevectors were considered, and linear combinations of degenerate unstable modes were fully explored in the multidimensional space only up to 2-fold degeneracy. We note that the energy scale of this problem is somewhat smaller than those previously reported for fluorides and oxides.\cite{Smith2015, Rahim2020} Care has been exercised to perform calculations with a well-converged basis set and $k$-mesh, and energy comparisons between phases with different unit cell sizes and shapes are made using equivalent $k$-meshes for maximal cancellation of errors in the discretization of Brillouin zone integrals. Subsequent to this extensive structure search procedure using an affordable GGA functional, the structural descriptions of phases I and II were improved utilizing a screened hybrid functional to reduce self-interaction error,\cite{Henderson2010} as described in the next section.

\begin{figure}
  \centering
    \includegraphics[width=\textwidth]{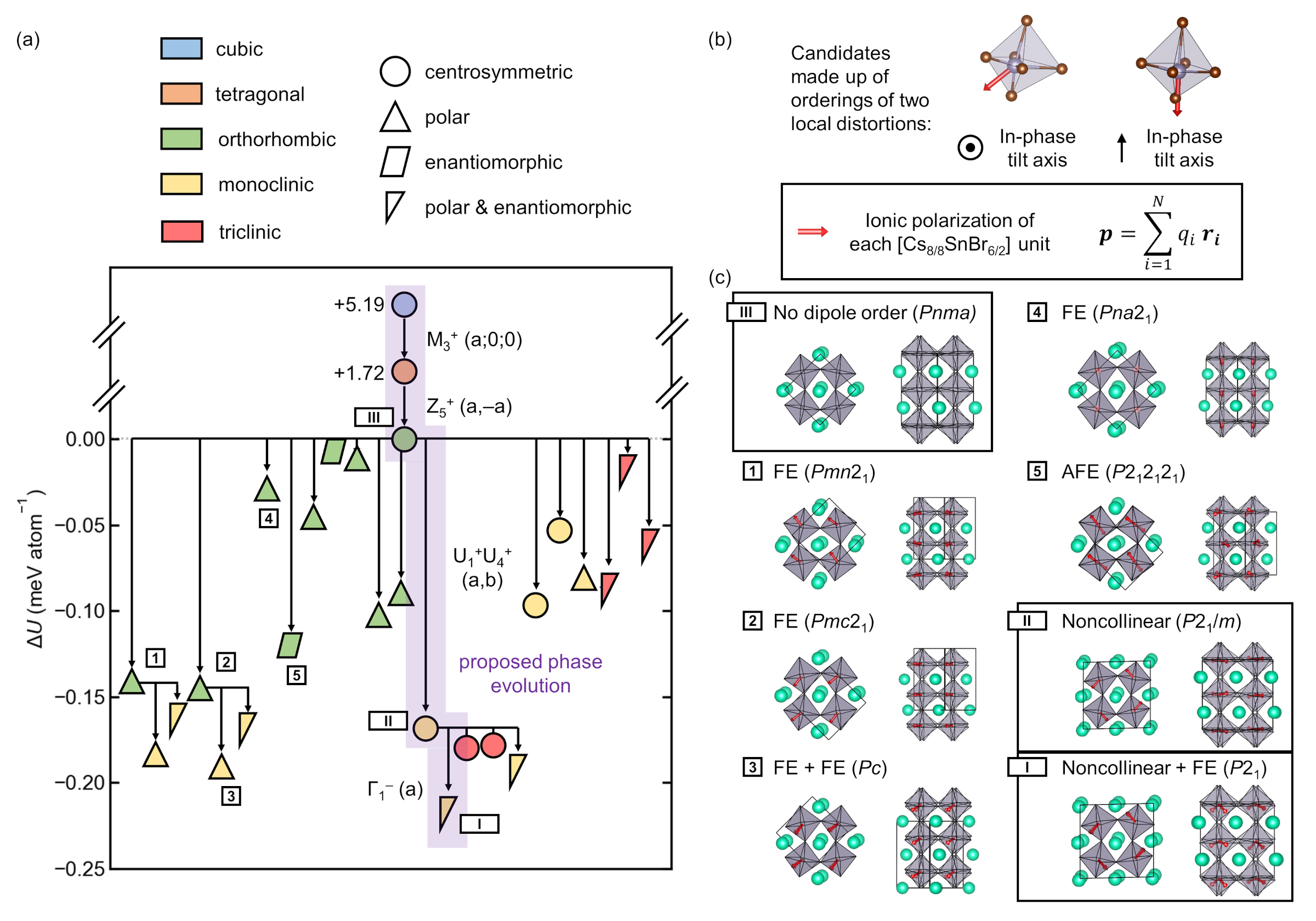}
  \caption{Solving the low temperatures structures of CsSnBr$_3$ by first-principles phonon mode mapping. (a) The tree of candidate phases, their energies, and their phase transition relationships. The crystal system of each phase is indicated by color, and the point group symmetry by shape. The proposed phase evolution is highlighted and the associated irreducible representations and order parameter directions are annotated. (b) All candidate structures for phases I and II are made up of different orderings of the same two local distortions. Stereochemical expression of the Sn(II) lone pair towards an octahedral edge or an octahedral vertex distorts both Sn and the Br ligands, creating a local electric dipole which is indicated by the red arrow (computed in the point charge limit). (c) Visualization of several structures, including that of phase III and those proposed for phases I and II. The ordering of local dipoles is described as well as the resulting space group symmetry (``FE'' = ferroelectric, ``AFE'' = antiferroelectric).}
  \label{Fig:Struc-Tree}
\end{figure}

Following these zero-temperature instabilities in the structure of phase III (see the 1D and 2D energy surfaces in Figure~S5) leads to 15 candidate structures for phase II, which are represented graphically in Figure~\ref{Fig:Struc-Tree}a. While these candidates exhibit a range of point group symmetries and unit cell sizes, they all comprise various orderings of two similar polar distortions of the Sn(II) coordination environment which allow for the Sn(II) lone pair to be stereochemically oriented towards an octahedral edge or an octahedral vertex (Figure~\ref{Fig:Struc-Tree}b). These distortions are conveniently visualized and discussed in terms of the local electric dipole they create. These electric dipoles order ferroically (parallel, ``FE''), antiferroically (anti-parallel, ``AFE''), or even in a noncollinear fashion. Additional candidates are visualized in Figure~S6 and include those which exhibit longer wavelength modulations and those with interactions between (individually) nonpolar tilting and nonpolar AFE electric dipole modes whose coupling results in imperfect dipole compensation and a net polarization (``FiE'', for ferrielectric). The latter appear to be a new manifestation of hybrid improper ferroelectricity.\cite{Benedek2011}

Three of the candidates (labeled ``1,'' ``2,'' and ``4'' in Figure~\ref{Fig:Struc-Tree}) are the phases enumerated by Stokes and coworkers which result from the combination of the $a^-b^+a^-$ tilting (in Glazer notation) of phase III with ferroic ordering of polar distortions along different directions.\cite{Stokes2002} While these structures, which have precedent among known perovskites,\cite{Johnston2010} were expected, an unexpected scenario is found to be more favorable. The proposed structure of phase II exhibits noncollinear electric dipole order of the distortions driven by the Sn(II) lone pair in the plane normal to the unique tilt axis ($b$ for phase III in the $Pnma$ setting of space group \#62). This results in a larger superstructure ($2\times2\times2$ with respect to the cubic aristotype) with space group symmetry $P2_1/m$ (\#11), preserving the centrosymmetry of phases III$-$V. While many perovskites are reported in the same space group, examination of unit cell sizes, Pearson symbols, and Wyckoff sequences reveals that this is indeed a new structure type (Table~S2).

We next enumerated the new minima or saddle points which result from following the unstable zone-center phonons of the three lowest energy candidates for phase II (labeled ``1,'' ``2,'' and ``II'') in Figure~\ref{Fig:Struc-Tree} (computed energy surfaces given in Figure~S7). This results in eight candidate structures for phase I. These candidates, including those labeled ``3'' and ``I'' and others in Figure~S6, combine the electric dipoles normal to the unique tilt axis, present in their parent phases, with additional dipoles along the unique tilt axis. The lowest energy candidate, which we propose as the correct structure of phase I, combines the noncollinear electric dipole order of its centrosymmetric parent phase with ferroic dipole order in the third dimension. This phase is polar (with polarization along the unique tilt axis) and chiral. This structure, a $2\times2\times2$ superstructure with respect to the cubic aristotype and with space group symmetry $P2_1$ (\#4), is also a new structure type (Table~S2). We note that the structures we propose for CsSnBr$_3$ phases I and II are distinct from the polar monoclinic phase recently proposed for CsSnBr$_3$ and CsSnI$_3$ by an alternative theoretical approach,\cite{Swift2023} which possesses mirror symmetry and a smaller unit cell and does not index the observed synchrotron diffraction (\textit{vide infra}).

The proposed complete phase evolution of CsSnBr$_3$ is highlighted in Figure~\ref{Fig:Struc-Tree}a. Phase III exhibits the familiar $a^-b^+a^-$ octahedral tilting common to a plethora of perovskites with undersized $A$-cations. The noncollinear electric dipole order which emerges in phase II and persists in combination with ferroic dipole order in phase I represents a new structural motif in perovskites. The irreducible representation of this noncollinear distortion is $U_1^+U_4^+$ with respect to orthorhombic phase III (when expressed in the $Pnma$ setting), or $X_5^-$ with respect to the cubic phase V, due to zone folding. Example transformations resulting from this irreducible representation and those considered by Stokes and coworkers\cite{Stokes2002} are given in Figure~S8. Additionally, a complete B{\"a}rnighausen tree\cite{Baernighausen1980} summarizing the symmetry relations\cite{Mueller2013} between phases I$-$V is presented in Figure~S9. The symmetry relations confirm that these transitions are compatible by Landau theory and renormalization group theory with the 2nd order transitions between phases I$-$IV assigned from the specific heat (Figure 1a). Notably, the index of the subgroup of phase I in the group of phase V is 192 (\textit{klassengleiche} index, $i_k$~=~8; \textit{translationengleiche} index, $i_t$~=~24), which appears to be the largest of any unsubstituted, commensurately modulated perovskite observed thus far. By this metric, CsSnBr$_3$ phase I exhibits the most complex (single) perovskite structure known other than incommensurate spin- or charge-density waves. Lastly, analysis of reported structures and phase transitions confirms that the finding of acentric Sn environments in CsSnBr$_3$ phases I and II is in line with chemical trends of lone pair stereochemical activity (Figure~S10) and leads to revision of the polar$-$nonpolar phase boundary among inorganic Cs$MX_3$ perovskites (Figure~S11).

The ground state of CsSnBr$_3$ can be contrasted with that of Cs$_2$AgBiBr$_6$, where a subtle, long wavelength modulation is reported from X-ray and neutron scattering.\cite{He2024} Some of the present authors find this Cs$_2$AgBiBr$_6$ phase to be nonpolar via dielectric spectroscopy,\cite{Cohen2022} suggesting distinct functionality from that of CsSnBr$_3$ phase I.

In general, systems which combine octahedral tilting with polar distortions of the $A$-cation environment, the octahedral cation environment, or both are exceedingly rare. Among halides, RbGeBr$_3$,\cite{Thiele1988} an elevated temperature phase of RbGeI$_3$\cite{Thiele1989}, and low temperature CsPbF$_3$\cite{Berastegui2001} exhibit this combination, resulting in structures which are polar but not chiral. Relative to CsSnBr$_3$, these phases all exhibit wider bandgaps and the Ge(II) compounds exhibit weaker spin-orbit interactions on the octahedral cation, presumably limiting the magnitude of bulk Rashba effects in these polar phases.

Noncollinear electric dipole order has attracted recent theoretical interest\cite{Lin2019, Zhao2020} but experimental observations are sparse.\cite{Yamada2019, Fukuda2020} In particular, Zhao and coworkers explored the possibility of an electric analog of the magnetic Dzyaloshinskii$-$Moriya interaction (DMI), finding that a one-to-one analogy exists and that the mechanism of the electric DMI does not rely on the spin-orbit interaction.\cite{Zhao2020} This accords with our finding that structures with noncollinear electric dipole order could be stabilized in DFT with scalar relativistic, rather than fully relativistic, treatment. Recently, a different noncollinear ordering of acentric Sn(II) distortions (with irreducible representation $X_5^+$) has been reported in a centrosymmetric phase of CH(NH$_2$)$_2$SnBr$_3$ (space group $Pa\bar{3}$) around room temperature by Yamada and coworkers.\cite{Yamada2019}

The II$-$I transition is a proper ferroelectric transition. The III$-$II transition destroys all mirror planes parallel to the rotation axis, resulting in a spontaneous toroidal moment and ferroaxial order.\cite{Johnson2011, Hlinka2016} The interaction of the noncollinear lone pair distortions with the extant octahedral tilts is essential to the ferroaxial order, possibly via imperfect cancellation of toroidal moments which otherwise sum to zero: The action of the noncollinear distortions alone on the untilted cubic aristotype does not produce ferroaxial order. Lastly, three of the four phase transitions are improper ferroelastic transitions: V$-$IV, with secondary order parameter $\Gamma_3^+$; IV$-$III, with secondary order parameter $\Gamma_2^+$; and III$-$II, with secondary order parameter $\Gamma_4^+$.

Thus, \textit{CsSnBr$_3$ is a ferroaxial$-$ferroelectric$-$ferroelastic multiferroic}. Extending the terminology from magnetic multiferroics,\cite{Khomskii2009} CsSnBr$_3$ is type I with respect to ferroaxial$-$ferroelectric coupling, and type II with respect to ferroaxial$-$ferroelastic coupling.

The unusual combination of polar and chiral symmetry with 3D connectivity and dispersive bands\cite{Huang2013} in this black, heavy semiconductor may lead to useful couplings between charge, spin, and polarized light in CsSnBr$_3$ phase I, a matter to which we will return after experimentally corroborating the proposed structures.

\textbf{Experimental confirmation of the proposed crystal structures of CsSnBr$_3$ phases I and II.} Here we demonstrate that a range of experimental observations, leveraging different physics, are all consistent with the unprecedented model for phase I identified by first-principles phonon mode mapping. Due to the narrow temperature range of phase II (77~K<$T$<85~K), we primarily corroborate its crystal structure indirectly via its relationship to phases I and III.

\begin{figure}
  \centering
    \includegraphics[width=\textwidth]{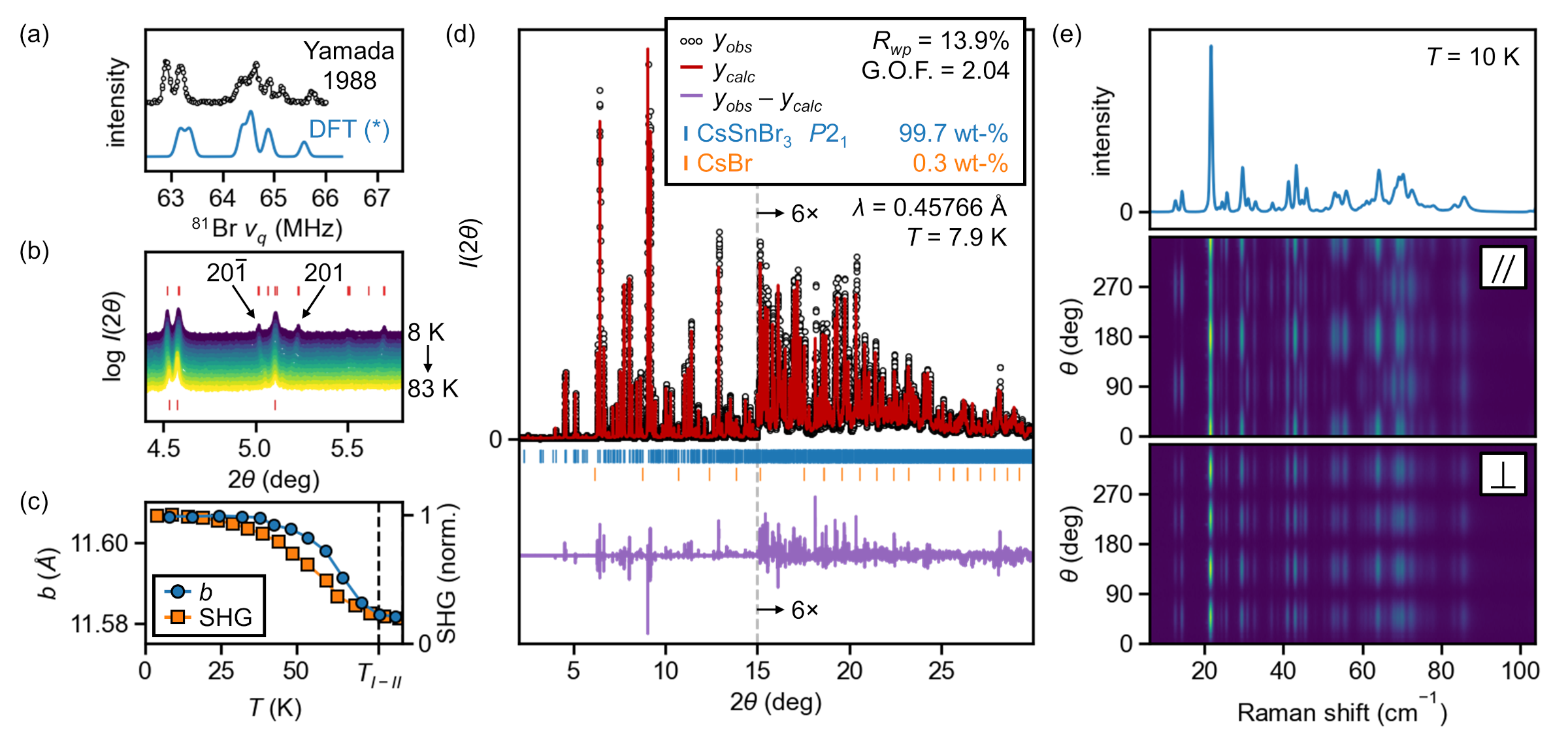}
  \caption{Experimental confirmation of the proposed structure of CsSnBr$_3$ phase I. (a) The experimental $^{81}$Br NQR spectrum at 77~K\cite{Yamada1988} strongly resembles that computed for the proposed structure (PBE, see also Figure~S12). (*) The DFT-computed spectrum has been shifted by $-7.4$~MHz to account for systematic error in field gradients or the tabulated quadrupole moment.\cite{Pyykko2018} Data from Yamada \textit{et al.} reprinted with permission from the Chemical Society of Japan.\cite{Yamada1988}
  (b) Excerpt of the temperature-evolution of synchrotron X-ray powder diffraction data from 8~K (purple) to 83~K (yellow) showing key superstructure Bragg reflections. Red ticks at the top are reflections from the proposed phase I structure model, and those at the bottom are the reflections from the structure of phase III. Notably, the $20\bar{1}$ and $201$ reflections (which do not correspond to reciprocal lattice points in phase III and whose splitting reflects the monoclinic angle) are seen to grow in intensities and split further on cooling (see also Figure~S14).
  (c) Pronounced elongation of the $b$ axis on cooling in phase I closely tracks the observed SHG signal (``norm.'' = normalized to maximum value), consistent with development of spontaneous polarization along [010] in the $P2_1/m$ to $P2_1$ transition ($T_{I-II}$).
  (d) Graphical results of Rietveld refinement at 7.9~K. See text.
  (e) Raman spectra at 10~K up to 100~cm$^{-1}$ Stokes shift. The unpolarized spectrum is in the top panel, and PO spectra (colors on a logarithmic scale) are given in the middle and bottom panels for incident and scattered beams in parallel and cross polarizations, respectively. See text.}
  \label{Fig:Diffraction-Raman}
\end{figure}

Electric field gradients at the nuclei, as probed by NQR and M{\"o}ssbauer spectroscopies, are sensitive probes of local symmetry. In the first paper which hints at an additional low temperature phase of CsSnBr$_3$, Yamada and coworkers observed that the two $^{81}$Br NQR signals corresponding to the distinct sites in phases III and IV split at 85$\pm$2~K into at least seven signals at 77~K which are not fully resolved,\cite{Yamada1988} consistent with our proposed model for phase II with eight distinct Br sites (Figure~S12a). This observation is incompatible with most other candidate structures for phase II from our phonon mode mapping as well as the recently proposed $Pc$ structure,\cite{Swift2023} which have fewer distinct Br sites. We computed the $^{81}$Br quadrupole resonance frequencies for our proposed structure models of phases I and II and 22 other models from DFT electric field gradients ($V_{zz}\approx230$~V~\AA$^{-2}$).\cite{Petrilli1998} We find excellent agreement for the model of phase I with the reported spin-echo-detected experimental spectrum at 77~K (Figure~\ref{Fig:Diffraction-Raman}a and Figure~S12a).\cite{Yamada1988} Agreement for the model of phase II is less favorable, suggesting persistent local distortions resembling the structure of phase I, \textit{i.e.} order-disorder character of the I$-$II transition. This scenario is reminiscent of the persistent pyramidal [GeBr$_3$]$^-$ environments detected by Raman spectroscopy in the cubic phase of CsGeBr$_3$.\cite{Gao2021} The electric field gradients at the Sn nuclei are substantially smaller ($V_{zz}\approx-10$~V~\AA$^{-2}$) than those at the Br nuclei leading to $^{119}$Sn M{\"o}ssbauer quadrupolar splittings ($\Delta E_Q\approx0.09$~mm~s$^{-1}$) consistent with the observed spectra (Figure~S3).

High-resolution synchrotron powder diffraction data are presented in Figure~\ref{Fig:Diffraction-Raman}b and \ref{Fig:Diffraction-Raman}d. Remarkably, aside from several very weak reflections we attribute to an unknown, trace hydrolysis product (\textit{vide infra}), the proposed model of phase I indexes the observed reflections once accounting for the cryostat window and trace CsBr. Similarly, all reflections expected from the model are detected, with the exception of those which are expected from the atom positions in the computational model to have intensities several orders of magnitude weaker than the strongest Bragg reflection, including 001 and 100. Notably, the plausible alternative models we tested, whether higher energy candidates from our first-principles phonon mode mapping, structures proposed by others,\cite{Swift2023} or previously reported perovskite structures with subgroups compatible with second-order transitions from phase III as implied by the heat capacity data, do not index the observed diffraction accurately (Figure~S13).

In comparison to phase III, phases I and II exhibit new reflections due to the larger superstructure and the loss of both glide planes and two of the three screw axes. Additionally, splitting of some Bragg peaks occurs due to the monoclinic distortion from orthorhombic symmetry. As a prominent example highlighted in Figure~\ref{Fig:Diffraction-Raman}b, the 20$\bar{1}$ and 201 peaks emerge which do not correspond to reciprocal lattice points in phase III (see Figure~S14), and their separation grows on cooling, reflecting greater deviation of the monoclinic angle from 90$^{\circ}$ at low temperature. As expected for $P2_1$ and $P2_1/m$, the Bragg reflection conditions are identical for all patterns recorded up to 83~K (\textit{e.g.}, Figure 3b). As the intensities of the new reflections are very weak in the narrow stability range of phase II, we focus our Rietveld analysis on phase I. Pawley refinement of the variable temperature data reveals pronounced elongation of the monoclinic $b$ lattice vector on cooling in phase I which closely tracks the observed SHG signal (Figure~\ref{Fig:Diffraction-Raman}c). This is reminiscent of the polar axis elongation observed in many ferroelectrics, and supports the proposed symmetry-breaking from $P2_1/m$ to $P2_1$ at the II$-$I phase transition.

The structures of phases I and II have the same unit cell size, the same Friedel symmetry ($2/m$), and the same systematic absences (due only to the $2_1$ screw axis). The deviations of atom positions in phase I ($P2_1$) from those of centrosymmetric phase II ($P2_1/m$) only manifest in the anomalous scattering contribution to the structure factors. As such, Rietveld refinements of the diffraction data at 7.9~K (details in the methods section) were performed using both space groups $P2_1/m$ (Figure~S15) and $P2_1$ (Figure~\ref{Fig:Diffraction-Raman}d). The refinements converged quickly, and multiple refinements with different sequences of parameter release led to the same minima. Crystallographic data and atomic positions for both models are provided in Tables~S3 and S4. The refinements led to acceptable G.O.F. values both for space group $P2_1/m$ (G.O.F.~=~2.06) and for space group $P2_1$ (G.O.F.~=~2.04) (Table~S3). However, the $R_{wp}$ values seem to be high (14.1~\% and 13.9~\%, respectively), and the refinements show some intensity misfits. This is a common feature for high-resolution synchrotron data and moreover, as the capillary had to be filled in air, the sample was only briefly ground in order to minimize the exposure to moisture and oxygen. This led to poor particle statistics as illustrated by artificial peak splitting (Figure~S16). This is also reflected in the comparatively high $R_{exp}$ value of 6.82~\%. Nevertheless, acceptable residual criteria like G.O.F. and $R_{F^2}$ and refined bond lengths in the expected range indicate reasonable refinement results.

A close inspection of the final Rietveld refinements reveals the presence of very small, unindexed reflections (Figure~S17). These cannot be assigned to any other Cs$-$Sn$-$Br phase, nor to any Sn$-$Br or other Cs$-$Br phase, nor to any oxide or hydroxide of tin or cesium or the two together, nor to water ice or any condensed gases from the capillary’s atmosphere, nor to any supercell with doubled or tripled $a$, $b$, or $c$ lattice parameters, nor to a removal of the 2$_1$ screw axis. Hence, we believe that these peaks are attributed to an unknown hydrolysis product of CsSnBr$_3$ formed by trace water from the atmosphere during grinding or emitted from the clay that was used for sealing.

We also tested the inclusion of  a second CsSnBr$_3$ phase with the structure of phase III ($Pnma$) usually found above 85~K, on the hypothesis that extended defects could inhibit the ferroelastic transition (Figure~S18). This naturally results in improved error metrics, but the resulting atom positions in phase I are negligibly affected and it is not conclusive if this added phase is physically justified or serves to collect intensity errors associated with poor particle statistics or anisotropic peak broadening.

Reduction of the space group symmetry from $P2_1/m$ to $P2_1$ leads to an increase in the number of independent parameters from 90 to 116 (Table~S3). As a result, an improvement of the fit is expected. The slight improvement of the G.O.F. from 2.06 to 2.04 does not prove the absence of the mirror plane. By reduction of the space group symmetry, atoms located on the mirror plane in the $P2_1/m$ model (Cs1-4 and Br5-8, Table~S4) obtain an additional degree of freedom. The Cs cations do not shift from the mirror plane and the Br anions show only minor displacements from their positions in the centrosymmetric crystal structure. Hence, the refined atom positions do not prove the absence of the mirror plane. The lower atomic displacement parameters for Sn and Br in the $P2_1$ structure (Table~S4) and the stability of the refinement give some very weak support for the $P2_1$ structure model. Thus, on its own, Rietveld refinement of powder diffraction data does not unambiguously prove the loss of inversion and mirror symmetries in phase I. However, the predictions of our first-principles mode mapping (Figure~\ref{Fig:Struc-Tree}), the favorable agreement between the computed and measured $^{81}$Br NQR spectrum (Figure~\ref{Fig:Diffraction-Raman}a and Figure~S12a), and the pronounced elongation along the 2$_1$ axis accompanied by the onset of significant SHG (Figure~\ref{Fig:Diffraction-Raman}c) strongly support the $P2_1$ model of phase I.

Polarized Raman spectroscopy additionally corroborates the proposed structure of phase I. The unpolarized Raman spectrum of a single crystal and the polarization$-$orientation (PO) dependence for parallel and cross polarizations, all at 10~K and up to 100~cm$^{-1}$ Stokes shift, are given in Figure~\ref{Fig:Diffraction-Raman}e (unpolarized temperature-dependence in Figure~S20). The clear angular periodicity of the PO data suggests a single crystal domain is probed.

For each spectrum corresponding to an orientation step of the PO data, we fitted intensities of pseudo-Voigt peaks while the centers, widths, and Gaussian/Lorentzian fractions were fixed based on initial fitting to the unpolarized spectrum (Figure~S21). The PO-dependence of these resulting intensities were then fit to the expected form of Raman modes of symmetry A or B, accounting for optical birefringence. Derivation of this PO-dependence for birefringent crystals in point group 2 ($C_2$) is given in the Supporting Information, Section~10. According to factor group analysis, the proposed structure of phase I should exhibit 120 Raman active modes, including 59 non-acoustic modes of A symmetry, and 58 non-acoustic modes of B symmetry (Table~S9).

At least 36 modes of A symmetry are robustly detected within the experimental range (Figure~S22, with indications of additional weak or unresolved modes), compared to 47 A modes calculated by density functional theory to lie within the experimentally accessed frequency range. No modes of B symmetry are experimentally detected, and B modes are expected to be silent for light propagation along the $b$ axis (see the Supporting Information, Section~10). Three different crystals were measured in backscattering geometry, with all showing no modes of B symmetry at base temperature. The monoclinic $b$ axis of phase I is not controlled by the orientation of the sample at room temperature, but rather by the direction that the 4-fold axis (tetragonal $c$) nucleates in the first-order phase V--phase IV transition on cooling, as this evolves through the continuous phase transitions to become the 2$_1$ screw axis in phase I. It is conceivable that the nucleation of the tetragonal $c$ axis is biased along the vertical (normal to the sample stage and parallel to the light propagation), for example by differences in surface energies for the exposed tetragonal \{110\} and \{001\} faces, because the macroscopic sample shape is square prismatic rather than cubic (the crystals do not grow on the face which rests on the bottom of the vial).

Thus, the PO data are compatible with the proposed structure of phase I assuming several additional A modes are too weak to be robustly fitted or too poorly resolved from others in frequency. Notably, these $\geq 36$ modes with A symmetry are incompatible with all considered alternative structure models, including higher energy candidates from our phonon mode mapping as well as structures proposed by others (Table~S9).

Previously reported $^{81}$Br NQR and our measured second harmonic generation, synchrotron powder diffraction, and polarized Raman spectroscopy are all consistent with the proposed new structures of CsSnBr$_3$ phases I and II identified by first-principles phonon mode mapping. While producing the same symmetries and resulting from distortions with the same irreducible representations, the geometric directions of the noncollinear lone pair distortions within the (010) plane are different (essentially by a 90$^{\circ}$ rotation about [010]) for the Rietveld-refined experimental models and the models computed via the DFT-PBE modemapping, implying a sensitivity of the computed models to cell size and shape or to the level of theory or implying that the experimental models are in a metastable, local minimum. To address this discrepancy and to improve our structure models generally (given the intrinsic challenges associated with displacements from centrosymmetry in the diffraction experiments and the known shortcomings of GGA-DFT in describing localized states)\cite{Henderson2010} we relaxed our structure models for phases I and II using a screened hybrid functional. For each phase ($P2_1$ and $P2_1/m$), structure models were relaxed using two different configurations to resolve the ambiguity above: 1) Fixed unit cell relaxation from the models Rietveld-refined against powder diffraction at 7.9~K, and 2) Fixed unit cell relaxation from models combining the experimental lattice parameters with the atom positions from the results of the DFT-only modemapping. Visualization of the Rietveld-refined experimental structures and comparison with the computed structures of phases I and II are presented in Figure~S23a. The computed $^{81}$Br NQR frequencies compare much more favorably to those from experiment for the models relaxed from the Rietveld-refined structures (Figure S23c), resolving the ambiguity about the direction of the distortions within the (010) plane.

The crystal structures Rietveld refined against X-ray diffraction data have been deposited in the Inorganic Crystal Structure Database (deposition numbers 2325515 and 2325516) and the crystal structures relaxed using a screened hybrid functional are included as Supporting Information files with technical parameters of the calculations included in the file headings.

\textbf{Physical properties of CsSnBr$_3$ related to electronic structure and polarization switching.} Aside from the heat capacity and nonlinear optical susceptibility presented in Figure~\ref{Fig:Cp-SHG} to establish the new phase transitions and the bulk noncentrosymmetry of phase I, we probed several properties related to the electronic structure and polarization. Although direct measurement of chiral optical responses is beyond the scope of this study, we note that $P2_1$ is the only noncentrosymmetric monoclinic space group with a $2_1$ axis, and thus the only one compatible with the crystal system and systematic absences from diffraction and the measured SHG response.

Unfortunately, the high bulk conductivity (\textit{vide infra}) of CsSnBr$_3$ samples precludes typical dielectric and ferroelectric characterization, including audio-frequency dielectric spectroscopy, polarization switching, and quantification of the spontaneous polarization, $\mathcal{P}_s$, via pyrocurrent integration. Instead, $\mathcal{P}_s$ was evaluated using the structure models relaxed using screened hybrid functionals. $\mathcal{P}_s$ was computed in the linear approximation, valid for small displacements from the centrosymmetric parent phase,\cite{Resta1993} as:

\begin{equation}
    \mathcal{P}_s = \frac{\left |e \right |}{\Omega}\sum_i{Z_i^* u_i},
\end{equation}

\noindent where $e$ is the electron charge, $\Omega$ is the unit cell volume, $Z_i^*$ are the Born effective charges of the ions in the parent phase, and $u_i$ are the displacements of the ions with respect to the parent phase. This yields $\mathcal{P}_s$~=~4.5~$\mu$C~cm$^{-2}$ parallel to the ferroic lone pair distortions along [010]. This value is several times that computed in the ionic point charge limit, reflecting the highly polarizable Sn$-$Br bonding environment associated with $s^2$ main-group metals,\cite{Du2010, Fabini2016, Fabini2020} nineteen times larger than that of Rochelle salt,\cite{Valasek1921} and eight times smaller than that of KNbO$_3$.\cite{Resta1993}.

In line with the expected influence of chemical pressure on pseudo-Jahn$-$Teller effects\cite{Bersuker2013} and with previous reports on thallium halides and main group halide perovskites,\cite{Du2010,Fabini2016} DFT calculations indicate that the acentric lone pair distortions in CsSnBr$_3$ should be enhanced significantly under tensile strain (Figure~S24). The long-range ordering of lone pairs cannot be readily predicted without a costly epitaxially-constrained first priniciples mode mapping, which is beyond the scope of this study. If the ordering remains that observed for bulk samples, one expects the spontaneous polarization and Rashba coefficients to be enhanced. Tantalizingly, others have grown CsSnBr$_3$ under several percent compressive strain,\cite{Wang2017, Wang2017-2, Zhang2021} suggesting similar magnitudes of tensile strain may be possible.

\begin{figure}
  \centering
    \includegraphics[width=\textwidth]{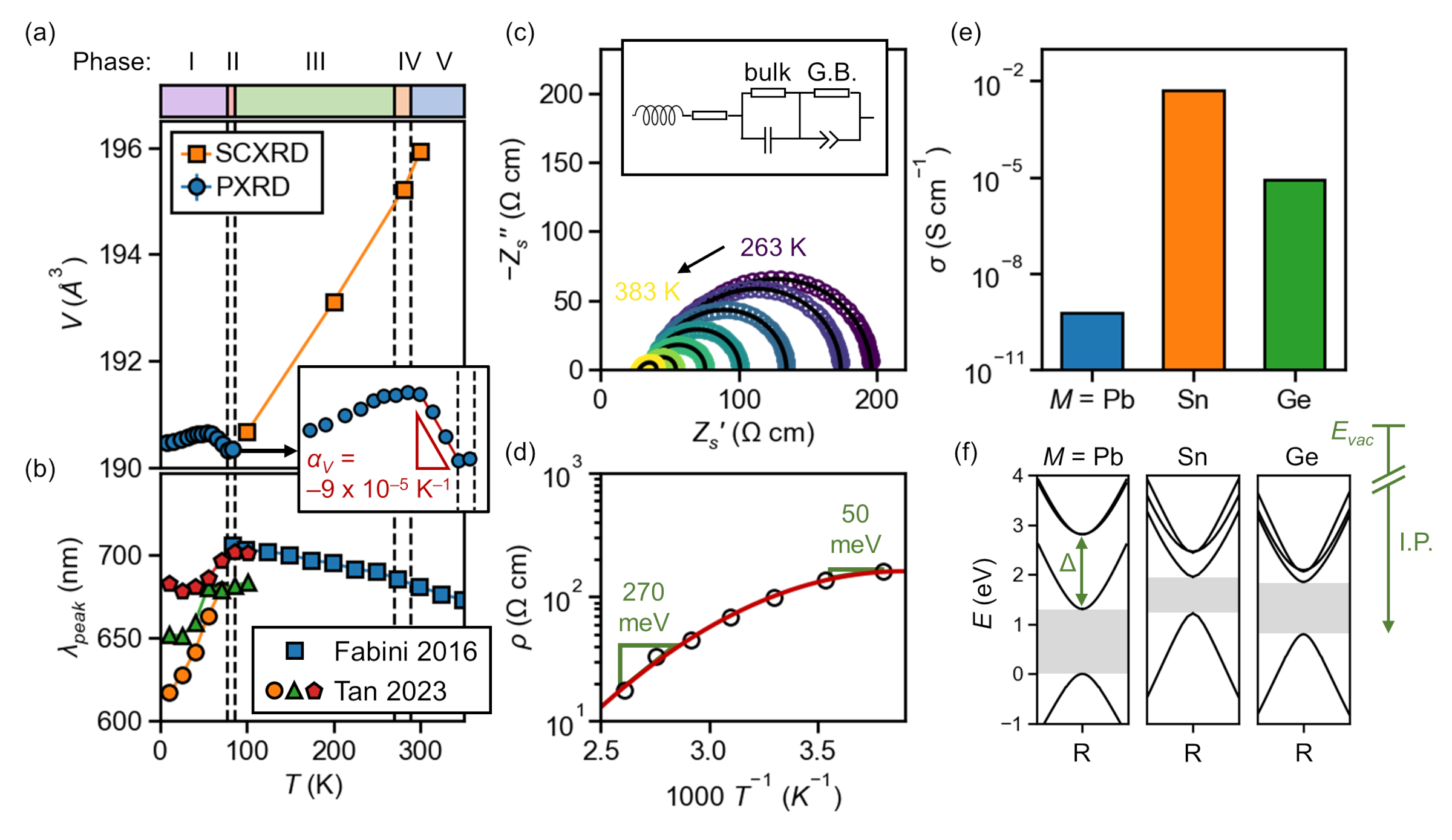}
  \caption{Physical properties of CsSnBr$_3$. (a) The unit cell volume, $V$, undergoes pronounced negative thermal expansion (NTE) with the onset of spontaneous polarization in phase I. (b) The reported PL in phase I is multi-peaked and blue-shifts dramatically on cooling,\cite{Tan2023} in stark contrast to the reported trend at higher temperatures,\cite{Fabini2016} offering clues about the band edge electronic structure. Data from Tan \textit{et el.} are reprinted with permission from the Editorial office of Chinese Physics B.\cite{Tan2023} (c) AC impedance spectra of polycrystalline CsSnBr$_3$ at various temperatures, fit with the inset model (G.B. = grain boundaries). (d) The total resistivity fit from the impedance spectra is not Arrhenius-like, but fitting activation energies near the high and low ends of the temperature range gives the small transport gaps annotated in the figure, suggesting shallow acceptors. (e) Conductivities of CsPbBr$_3$, CsSnBr$_3$, and CsGeBr$_3$ (spectra in Figure~S31). (f) Computed band structures (HSE06+SOC) of cubic Cs$M$Br$_3$ along $\lambda-$R$-\lambda$, aligned by core states to a common scale. Bandgaps are shown by gray shading. The ionization potentials (I.P.) strongly track the experimental conductivity, suggesting the electron chemical potential term plays a dominant role in the formation energy of $p$-type defects.\cite{Zhang1991} See text. $\Delta$ indicates the spin-orbit splitting of the outer shell $p_{1/2}$ and $p_{3/2}$ states of the group 14 metal.}
  \label{Fig:Properties}
\end{figure}

Negative thermal expansion (NTE) occurs in some ferroelectric phases, and the spontaneous polarization is known to be correlated with the magnitude of the NTE in PbTiO$_3$-related solid solutions,\cite{Chen2011} likely via an electrostrictive mechanism. The temperature evolution of the CsSnBr$_3$ unit cell volume from synchrotron powder diffraction is shown in Figure~\ref{Fig:Properties}a (lattice parameter evolution and a comparison with volumes from theory are given in Figures~S25 and S26). Pronounced negative thermal expansion, driven primarily by the polar $b$ axis (Figure~S25, $\alpha_b=\frac{1}{b}\left(\frac{\partial b}{\partial T}\right)=-7.8\times10^{-5}$~K$^{-1}$ over the window 60~K$<T<$77~K) is observed just below the transition to phase I. The volumetric thermal expansion coefficient, $\alpha_V=\frac{1}{V}\left(\frac{\partial V}{\partial T}\right)=-9.2\times10^{-5}$~K$^{-1}$ over the same window, is substantially more negative than for some ferroelectric oxides\cite{Chen2011} and roughly twice that of Sn$_2$P$_2$S$_6$,\cite{Rong2016} albeit over a narrower temperature range. It appears that $\alpha_V$ for particular oxide solid solutions may be more negative than the value for CsSnBr$_3$, but few temperature points are available for comparison.\cite{Pan2017} Along the polar $b$ axis, the effective polarization$-$electrostriction coefficient ($Q_{22}\approx$~1.1~m$^4$~C$^{-2}$ in Voigt notation) calculated from the DFT-computed polarization and experimental lattice expansion is on par with that of polyvinylidene difluoride,\cite{Zhang2021-JACS} smaller than those of several layered and 0-D hybrid main group halides,\cite{Wang2019, Zhang2021-JACS, Tao2023} and substantially larger than those reported for ferroelectric ceramics.\cite{Zhang2021-JACS} Among inorganic compounds, this value is exceeded by oxygen-deficient ceramics with slow electrostrictive responses,\cite{Yu2022} but the intrinsic mechanism at work in CsSnBr$_3$ suggests faster sensing or actuation should be possible in this material. Chen and coworkers have previously measured large electrostriction for centrosymmetric CH$_3$NH$_3$PbI$_3$ and ruled out an intrinsic mechanism in that compound.\cite{Chen2018}

Band-to-band photoluminescence (PL) or its absence should provide clues about the band-edge momentum- and spin-splitting induced by the spin-orbit interaction and broken inversion symmetry in phase I. The recently reported PL at low temperatures (Figure~\ref{Fig:Properties}b)\cite{Tan2023} shows a marked departure from the trend reported in phases III, IV, and V,\cite{Fabini2016} and agrees broadly with our own preliminary measurements in this regime. In phase I, three features are observed, with the higher energy emission attributed to free excitons and the two others to bound excitons, based on peak widths and excitation power studies.\cite{Tan2023} The peak positions of all three features blue-shift substantially on cooling into phases I and II, with the free exciton peak blue-shifting nearly 200~meV down to the lowest temperature measured. This shift is much larger than the blue-shift observed at some tilting transitions in hybrid lead bromide and iodide perovskites,\cite{Wright2016} which also occur only over a limited temperature window below the transitions.

This observation of bright photoluminescence attributed to free excitons\cite{Tan2023} from CsSnBr$_3$ phase I places constraints on the electronic structure of this phase.\cite{Steele2019, Lafalce2022} It was proposed that strong Rashba effects in both the valence and conduction bands of Pb(II) halide perovskites\cite{Kim2014} would suppress radiative recombination by causing direct transitions to be spin-forbidden.\cite{Zheng2015} Though such strong Rashba splitting of the valence band is not supported experimentally in Pb(II) bromide perovskites,\cite{Frohna2018, Sajedi2020} the same logic would ostensibly require that the bulk Rashba effect in CsSnBr$_3$ is small. We qualitatively tested the individual influences of thermal expansion, octahedral tilting, ferroic dipole order, and noncollinear dipole order on the band structure of CsSnBr$_3$ with and without the spin-orbit interaction (Figure~S27). We find that expansion and tilting preserve a direct bandgap, ferroic dipole order causes Rashba effects of differing magnitudes at both band edges with spin-orbit coupling (SOC), and noncollinear dipole order produces an indirect gap even without SOC. Certainly, the precise electronic structure of this phase, which combines all of these features to varying degrees, requires further study, including symmetry-based analysis of possible relativistic effects on the energetic ordering of excitonic states.\cite{Becker2018}

Polarization switching is desirable both for traditional ferroelectric applications and as a means of switching the spin texture in a bulk Rashba semiconductor. However, preliminary dielectric spectroscopy and magnetocapacitance measurements revealed CsSnBr$_3$ to be a leaky dielectric. We investigated this undesirably high conductivity by temperature-dependent impedance spectra (Figure~\ref{Fig:Properties}c), which reveal a very low total resistivity ($\sim$10$^2~\Omega$~cm) for a semiconductor with a gap of $\sim$1.9~eV.\cite{Li2018} The temperature-dependence of the resistivity (Figure~\ref{Fig:Properties}d) does not neatly follow Arrhenius behavior, suggesting that in addition to changing the thermal occupation of non-localized states, other factors are at play. These could include a thermally-aided spatial redistribution of point defects and, thus, the electronic carrier density. Indeed, solid state $^{119}$Sn nuclear magnetic resonance spectroscopy (NMR) suggests spatial redistribution of electronic carriers, likely from mobile point defects, at room temperature over long timescales (see Section~16 and Figure~S28 of the Supporting Information). Nonetheless, fitting the observed resistivities to an Arrhenius law gives activation energies substantially smaller than half the bandgap ($\sim$50~meV at 273~K, $\sim$270~meV at 373~K), consistent with shallow acceptors. We note that there is variation in the reported conductivity of CsSnBr$_3$ and even the sign of its temperature dependence, with our samples being among the most insulating, and hence, pure, thus far.\cite{Clark1981, Yamada1990, Xie2020, Zhang2020} Greater understanding and control of the point defects in CsSnBr$_3$\cite{Liang2021} may be necessary for some applications.

For comparison, we prepared and measured CsGeBr$_3$ and CsPbBr$_3$ (Figures~S29, S30, and S31), and their resistivities are plotted in Figure~\ref{Fig:Properties}e. CsPbBr$_3$ is many orders of magnitude more insulating ($\sim$10$^9$~$\Omega$~cm, in agreement with the value previously reported for samples made from commercial-grade precursors),\cite{Stoumpos2013} and CsGeBr$_3$ is intermediate ($\sim$10$^5$~$\Omega$~cm). We hypothesize that this trend reflects primarily a wide variation in $p$-type carrier density driven indirectly (\textit{vide infra}) by the relative energies and spatial overlap of the group 14 outer $s$ orbital with the Br 4$p$ orbitals. The hybrid functional (HSE06+SOC) band structures of cubic CsPbBr$_3$, CsSnBr$_3$, and CsGeBr$_3$ are shown along $\lambda-$R$-\lambda$ in Figure~\ref{Fig:Properties}f, aligned to a common energy scale by core states. Pb 6$s$ is relativistically contracted leading to a deep valence band center and the narrowest valence band width of the series. In contrast, the shallow Sn 5$s$ orbital is strongly interacting with the 4$p$ orbitals of the Br ligands, raising the valence band center and widening the bandwidth significantly. This leads to a shallower valence band maximum\cite{Li2019} which contributes to the high concentration of holes ($\sim10^{16}$~cm$^{-3}$ to $\sim10^{17}$~cm$^{-3}$)\cite{Zhang2020} via the influence of the electron chemical potential term in the defect formation energy.\cite{Zhang1991} The situation in CsGeBr$_3$ is intermediate, where the large relative size of Cs on the $A$-site and the less diffuse Ge 4$s$ orbital limits the spatial overlap between Ge 4$s$ and Br 4$p$, leading to a valence band maximum which lies between those of the Sn and Pb cases and a moderate hole concentration.

The intrinsically small ionization potential for bulk CsSnBr$_3$ suggests suppressing the formation of $p$-type defects may be challenging. Our own preliminary attempts utilizing hypophosphorous acid as a reducing agent during sample preparation were able to suppress the conductivity only by roughly a factor of two. Compensation doping, purification of precursors, use of alternative reducing agents, or other strategies\cite{He2023} may prove more effective.

The HSE06+SOC computed spin-orbit splittings of the group 14 metal $np_{1/2}$ and $np_{3/2}$ states for cubic Cs$M$Br$_3$ are 1.50~eV, 0.49~eV, and 0.22~eV at the R point for Pb, Sn, and Ge, as shown in Figure~\ref{Fig:Properties}f. This suggests that though Sn is substantially lighter than Pb, there is potential for a large Rashba effect in the conduction band if the polar symmetry-breaking is sufficiently large, and if the difference in on-site energies\cite{MeraAcosta2020} is not too great.

\section{Conclusion}

We have shown that CsSnBr$_3$ exhibits two new structure types at low temperatures, including a polar and chiral structure in phase I ($T<$77~K). The symmetry relations between phases I$-$V reveal CsSnBr$_3$ to be a ferroaxial$-$ferroelastic$-$ferroelectric multiferroic, allowing a rich variety of useful couplings between mechanical, electronic, and optical degrees of freedom. CsSnBr$_3$ phase I is the heaviest inorganic non-fluoride polar main-group halide perovskite known thus far and joins the small number of 3D polar perovskite halides including CsGe$X_3$,\cite{Thiele1987} RbGeBr$_3$,\cite{Thiele1988} CH(NH$_2$)$_2$Ge$_{0.5}$Sn$_{0.5}$Br$_3$,\cite{Liu2022} (MDABCO)NH$_4$I$_3$ (MDABCO = $N$-methyl-$N^\prime$-diazabicyclo[2.2.2]octonium),\cite{Ye2018} low temperature CH$_3$NH$_3$SnBr$_3$,\cite{Swainson2010} and low temperature CsPbF$_3$.\cite{Berastegui2001} CsSnBr$_3$ has a narrower bandgap than all of these, heavier atoms contributing to the band edges than in CsGe$X_3$, RbGeBr$_3$, or (MDABCO)NH$_4$I$_3$, and much more dispersive bands than (MDABCO)NH$_4$I$_3$, all of which may contribute to the strength of possible bulk Rashba effects at the band edges.

Partial, isovalent chemical substitution on all three sites also offers an opportunity to enhance possible bulk Rashba effects: larger $A$-cations, Ge on the octahedral site, or Cl on the anion site should all enhance the symmetry-breaking, while Pb on the octahedral site or I on the anion site should enhance the spin-orbit interaction. The optimal tradeoff between these competing factors and the possibly deleterious impacts of alloy disorder are not yet known.

Taken together, the polar crystal structure, substantial spin-orbit splitting of the Sn 5$p$ orbitals, narrow bandgap, and sensitivity of lone pair distortions to tensile strain suggest bulk samples and epitaxial films of CsSnBr$_3$ or its neighboring solid solutions as strong candidates for bulk Rashba semiconductors. Direct investigation of the electronic structures and spin textures of such materials by relativistic first-principles methods, transport measurements, or photoemission spectroscopy presents a potentially fruitful area for future study.

The discovery of new phases of interesting symmetry 50 years after the first preparation of CsSnBr$_3$ and over a decade into intensive study of main group perovskite halides for optoelectronics suggests that more surprises may await in this curious and promising family of materials.

\section{Experimental and Computational Methods}

\textit{Sample preparation}: CsSnBr$_3$ for synchrotron powder diffraction was prepared in Chicago by Bridgman crystal growth. Stoichiometric amounts of CsBr (Sigma Aldrich, 99.999\%, 36~mmol scale) and SnBr$_2$ (Sigma Aldrich, 36~mmol) were mixed and flame-sealed in a fused silica ampoule under vacuum. This ampoule was reacted in a box furnace, with 10~h ramp to 650$^{\circ}$~C, holding for 10~h, and cooling in 10~h. The resulting ingot was removed from the tube and the surfaces were scraped clean as some impurities were observed to be stuck to the glass. The cleaned ingot was broken into chunks and flame-sealed in a sharp-tipped ampoule under vacuum for Bridgman crystal growth. Bridgman growth was conducted with a hot zone temperature of 775$^{\circ}$~C and no cold zone temperature – this produces the largest temperature gradient. The ampoule was heated for 7~h in the hot zone, then traversed downwards at a rate of 10~mm~h$^{-1}$. The resulting shiny black ingot was cut and polished to a shiny optical finish. Powder for the synchrotron XRD was obtained by isolating a chunk of this polished material and powdering it with mortar and pestle just prior to XRD experiments.

CsSnBr$_3$ for all other experiments was prepared in Stuttgart by three different routes:

1) Polycrystalline ingots were prepared by solidification from the melt of binary bromides at the 10~mmol scale. Stoichiometric quantities of CsBr and SnBr$_2$ powders were pressed into a pellet in an argon glovebox and loaded into a fused silica ampoule, which was flame-sealed under vacuum. The ampoule was heated to 600$^{\circ}$~C over 12~h, held for 6~h, and then cooled to ambient temperature over 12~hours. This yields a monolithic, black ingot which is easily separated from the ampoule wall.

2) Single crystals were grown by slow cooling from ethylene glycol solution of the binary bromides. CsBr and SnBr$_2$ powders (with SnBr$_2$ in 10\% molar excess, typical scale 1~mmol) were loaded into $\sim$1~cm diameter screw top vials and capped with silicone septa in an argon glovebox. Before use, the round bottoms of the vials were gently flattened in a flame to prevent crystal nuclei from sliding under gravity to the bottom and intergrowing in one mass, and the vials were cleaned in a base bath. Outside the glovebox, dry ethylene glycol (typical concentration 1.5~ml~mmol$^{-1}$ of product; the combination of scale and concentration depends on the area of the growth vessel bottom face to maximize crystal size without intergrowth of adjacent nuclei) was added using Schlenk technique and the vials were heated in a metal heating block on a hotplate to 150$^{\circ}$~C, yielding a transparent, yellow solution. The vials were slow-cooled on a programmable hotplate (Heidolph Instruments) from 125$^{\circ}$~C to 25$^{\circ}$~C over 99~h. The products were transferred under flowing argon into a Schlenk-frit, washed thrice with 5~ml dry ethanol, dried under vacuum, and transferred into the glovebox. This procedure yields black, rectilinear crystals (\{100\} facets) of CsSnBr$_3$ up to 7 mm on a side which are squat in the vertical direction (no growth on the bottom surface resting against the vial), as well as trace quantities of physically separate red truncated octahedra which appear to be Cs$_2$Sn$_{1+\delta}$Br$_6$. It is unknown whether the trace oxidation of Sn(II) to Sn(IV) occurs during the extended crystal growth procedure due to imperfect seals, the quality of the argon feed gas, or the purity of the solvent.

3) Preliminary experiments to reduce the unintentional hole doping were pursued by precipitation from hydrobromic acid in the presence of strongly reducing hypophosphorous acid. First, 762.4~mg (2.34~mmol) Cs$_2$CO$_3$ was reacted with a 2.34~ml of a 1:1 v/v mixture of 48\% HBr and 50\% H$_3$PO$_2$ in a round bottom flask in air, evolving CO$_2$. This flask was connected to the Schlenk line and sparged copiously with argon for several hours. Separately, 1303~mg (4.68~mmol) SnBr$_2$ was dissolved under argon in 4.68~ml of the same acid mixture upon stirring at 80$^{\circ}$~C, yielding a transparent, yellow solution. The Cs$^+$ solution was transferred by cannula into the flask containing the Sn$^{2+}$ solution, yielding an immediate black precipitate. After stirring for several minutes, the solution was cooled to room temperature. The product was washed with dry ethanol on a Schlenk frit, dried under vacuum, and transferred into the glovebox.

CsPbBr$_3$ was prepared by dissolving 2~mmol PbO in 2~ml hot, concentrated HBr under stirring in air. After dissolution, 1~mmol Cs$_2$CO$_3$ was gradually added, yielding an immediate bright orange precipitate. Stirring was ceased after an hour, the solution was cooled to room temperature, and excess solution was decanted. The remaining mixture was evaporated to dryness, yielding a bright orange powder which was easily ground and stored in dry air.

CsGeBr$_3$ was prepared from a 2:1~v/v mixture of 48\% HBr and 50\% H$_3$PO$_2$. All manipulations were performed under argon with Schlenk technique. 30~ml of the acid mixture was degassed, then transferred into a round bottom flask containing 300~mg GeO$_2$ (2.87~mmol). After complete dissolution and reduction of Ge(IV) to Ge(II) upon stirring for 1~h at 120$^{\circ}$~C in an oil bath, this solution was transferred hot to a flask containing 610.7~mg CsBr (2.87~mmol), yielding an immediate yellow-orange precipitate. This mixture was heated in an oil bath, dissolving to form a colorless solution, and then left to cool uncontrolled without stirring. The resulting several-mm yellow-orange needles were collected and washed on a Schlenk frit with dry ethanol.

\textit{Laboratory powder X-ray diffraction}: Powder X-ray diffraction at room temperature was recorded using a STOE Stadi-P diffractometer with Cu-K$_{\alpha1}$ radiation and a Ge(111) monochromator in STOE geometry, which combines aspects of the Debye-Scherrer and Guinier geometries. Samples were packed in borosilicate glass capillaries and rotated during acquisition.

\textit{Heat capacity}: Measurements on a single crystal (16.07(1) mg) and a flat shard of a polycrystalline ingot (16.696(5) mg) were carried out using relaxation calorimetry in a Quantum Design PPMS between 1.8~K and 300~K. Poor thermal transport through the vertical thickness of the macroscopic single crystal restricted the region of strong thermal coupling to <200~K.

\textit{Nonlinear optical spectroscopy}: The second harmonic generation (SHG) measurements were performed using femtosecond optical pulses ($\tau_{lp}\approx250$~fs, central wavelength $\lambda_{lp}\approx800$~nm) from a Ti:sapphire amplifier (Coherent RegA 9000) with a repitition rate of 100~kHz seeded by a Ti:sapphire oscillator (Coherent Mira 900). The laser light was collimated with a telescope and power attenuated to 4~mW. The laser light was guided through a long pass filter and a dichroic mirror (DM1) and focused onto the sample with a 50x microscope objective (Newport) at normal incidence to a spot size of $\approx50$~$\mu$m. The sample, a crystal of several mm, was cleaved with a razor blade and glued to a sample holder using GE varnish, then mounted in a Cryovac Konti continuous flow cryostat. The generated SHG signals (400~nm) were reflected from DM1 and collected by a photomultiplier tube (Hamamatsu H10720-113) after being short-pass filtered. The signals were further processed with a lock-in amplifier. The SHG intensity was measured at temperatures between 4~K and 110~K.

\textit{M{\"o}ssbauer spectroscopy:} The $^{119}$Sn M{\"o}ssbauer spectroscopic study on the CsSnBr$_3$ sample was performed with a Ba$^{119m}$SnO$_3$ source. The measurement was conducted in a continuous flow cryostat system (Janis Research Company LLC) at temperatures varying between 95~K and 6~K while the source was kept at room temperature. The sample was ground to a fine powder and mixed with $\alpha$-quartz to ensure an even distribution of the sample within the poly(methyl methacrylate) sample holder (diameter 2~cm). The optimal sample thickness was calculated according to Long \textit{et al.}\cite{Long1983} The program WinNormos for Igor was used to fit the spectrum.\cite{Brand2017}

\textit{Laboratory single crystal X-ray diffraction}: Crystals suitable for single-crystal X-ray diffraction (SCXRD) were selected under high viscosity oil, and mounted with some grease on a loop made of Kapton foil (Micromounts$^{TM}$, MiTeGen, Ithaca, NY). Diffraction data were collected between 50~K and 300~K with a SMART APEX-II CCD X-ray diffractometer, using graphite-monochromated Mo-K$\alpha$ radiation (Bruker AXS, Karlsruhe, Germany), equipped with a N-Helix low-temperature device (Oxford Cryosystems, Oxford, United Kingdom). The reflection intensities were integrated with the SAINT subprogram in the Bruker Suite software,\cite{Bruker2015} a multi-scan absorption correction was applied using SADABS\cite{Sheldrick2016} or TWINABS\cite{Sheldrick2012} and the structures were solved by direct methods and refined by full-matrix least-square fitting with the SHELXTL software package.\cite{Sheldrick2007, Sheldrick2015}

\textit{First-principles modeling}: All first-principles modeling was performed using the Vienna Ab initio Simulation Package (VASP), which implements the Kohn-Sham formulation of density functional theory using the projector augmented wave (PAW) formalism.\cite{Kresse1993, Kresse1996, Kresse1996-2, Kresse1999, Bloechl1994} For Cs, the 5$s$ and 5$p$ electrons were included in the valence, as were the 3$d$, 4$d$, and 5$d$ for Ge, Sn, and Pb, respectively. The plane wave basis set cutoff energy (564~eV) and reciprocal space mesh densities ($\sim$0.11~\AA$^{-1}$, corresponding to $\sim$5000~$k$-points per reciprocal atom), were chosen based on convergence of the total energy to better than 10$^{-4}$~eV~atom$^{-1}$. Forces were converged to better than 10$^{-5}$~eV~\AA$^{-1}$ before phonon calculations.

As discussed in the main text, the influences of exchange-correlation functional and dispersion corrections were studied on the harmonic phonons of CsSnBr$_3$ phase III, including the PBE functional\cite{Perdew1996} with and without DFT-D3 dispersion corrections,\cite{Grimme2010} and the PBEsol functional.\cite{Perdew2008} Subsequent relaxations, finite-displacement phonons, electric field gradients, and perturbation theory calculations (for ion-clamped static dielectric tensors and Born effective charges)\cite{Baroni1986, Gonze1997, Gajdos2006} utilized PBE without dispersion corrections, the only choice which correctly found 0~K mechanical instability of phase III. Electric field gradients were computed by the PAW method of Petrilli and coworkers\cite{Petrilli1998} and converted to $^{81}$Br nuclear quadrupole resonance frequencies and $^{119}$Sn M{\"o}ssbauer quadrupolar splittings using the updated nuclear quadrupole moments from Pyykk{\"o} ($^{81}$Br: $Q=+257.9$~mb; $^{119}$Sn: $Q=-132$~mb)\cite{Pyykko2018} and $^{119m}$Sn $\gamma$-ray emission of 23.870~keV.

Harmonic phonon dispersions were computed by the method of finite displacements\cite{Parlinski1997} with phonopy.\cite{Togo2023, Togo2023-2} The non-analytic correction for insulators\cite{Pick1970} was included in plotting phonon dispersions and was omitted in mapping the potential energies along (linear combinations of) phonon eigenvectors. Distorted structures were generated using ModeMap.\cite{Skelton2016} The small energy differences between relaxed candidate structures with different unit cell shapes or sizes were evaluated on equivalent $k$-meshes to maximize error cancellation in going from continuous integrals to discrete summations.

Subsequent to the first-principles modemapping using PBE, screened hybrid density functional calculations to reduce self-interaction error\cite{Henderson2010} and improve the structural description of phases I and II were performed using the HSE06 functional (25\% Fock exchange at short range, 0.2~\AA$^{-1}$ range separation, PBE for the DFT part)\cite{Heyd2003} on a reduced 3$\times$3$\times$3 $k$-point mesh with a reduced 362~eV plane wave basis set cutoff energy. Crystal structures were relaxed until force convergence was better than 20~meV~\AA$^{-1}$. For each phase, two configurations were considered: 1) Fixed unit cell relaxations with the experimental structures from PXRD as the starting guesses, and 2) Fixed unit cell relaxations from starting guesses which combine the experimental lattice parameters with the atom positions from the DFT-only modemapping results. Electric field gradients for these models were also computed at the HSE06 level for the NQR comparison in Figure~S23. Born effective charges for the linear approximation to the spontaneous polarization were computed at the PBE level using perturbation theory as parallelization constraints on VASP's finite field implementation (the only possibility for hybrid functionals) led to single-core memory requirements exceeding accessible resources for this large, low symmetry system.

The electronic structures of cubic CsGeBr$_3$, CsSnBr$_3$, and CsPbBr$_3$ were computed with a screened hybrid functional and spin-orbit coupling, HSE06+SOC,\cite{Heyd2003} with a reduced $k$-mesh density (7$\times$7$\times$7) and cutoff energy (465~eV).

Analysis of results made extensive use of ISODISTORT (H. T. Stokes, D. M. Hatch, and B. J. Campbell, ISODISTORT, ISOTROPY Software Suite, iso.byu.edu)\cite{Campbell2006} and the Bilbao Crystallographic Server.\cite{Aroyo2006,Aroyo2006-2,Aroyo2011} Crystal structures were visualized with VESTA.\cite{Momma2011}

\textit{Synchrotron diffraction}: High-resolution powder diffraction was performed on the 11-BM diffractometer at the Advanced Photon Source at Argonne National Laboratory, using a calibrated wavelength of 0.457660~\AA~and a closed flow helium cryostat (Oxford Instruments). An interior portion of a Bridgman-grown single crystal was ground briefly in air, packed in a Kapton capillary, and sealed with clay. Once inside the cryostat, the atmosphere was purged several times with helium. Due to the twelve-analyzer design,\cite{Lee2008} high-resolution data could be rapidly acquired ($\sim$20~minute scans). Diffraction was recorded at 287~K, then the sample was cooled to base temperature (7.9~K) over 50~minutes. Diffraction was recorded while dwelling at this temperature, then 12 patterns were recorded while ramping from base temperature to 82.6~K over 240~minutes, then a final pattern was recorded while dwelling at this temperature. The pattern recorded during the base temperature dwell and the first pattern during the up-ramp are indistinguishable, implying the sample temperature reached steady-state. Comparable experiments above 90~K have been reported previously.\cite{Fabini2016} Variable temperature Pawley refinements were performed using GSAS-II.\cite{Toby2013} TOPAS 6.0 was used for Rietveld refinement.\cite{Coelho2018} The background was modeled by a Chebychev polynomial of 6th order. The computed monoclinic structures were used as the starting model for the fully weighted Rietveld refinement\cite{Rietveld1969} using both space groups $P2_1/m$ (\#11) and $P2_1$ (\#4). As we detected some anisotropic peak broadening, indicative of structural disorder, we applied symmetry adapted spherical harmonics of 4th order. We also tested the omission of symmetry adapted spherical harmonics and the Stephens model (Figure~S19) which led to negligible changes in atom positions.

\textit{Raman spectroscopy}: The Raman scattering measurements were conducted using a home-built system in the back-scattering geometry. A continuous wave (CW) diode laser with a wavelength of 1.57~eV (Toptica Inc., USA) served as the light source. The laser beam was focused using a 50x objective (Zeiss, USA). The Rayleigh scattering was filtered out using notch filters (Ondax Inc., USA), and the scattered beam was focused into a 1~m spectrometer (FHR 1000, Horiba) with a 1800~gr~mm$^{-1}$ grating. The beam was captured by a CCD detector.
For the polarization orientation (PO) measurements, the incident beam underwent filtration using a polarizer and a half-wave plate (Thorlabs, USA). The scattered beam underwent further filtration using another half-wave plate oriented parallel or perpendicular to the incident beam's polarization, along with an additional polarizer. The beam underwent stepwise rotation by a half-wave plate placed in between the two polarizers.
The samples were cooled by a Janis cryostat ST-500 controlled by Lakeshore model 335 with liquid helium flow to reach 10~K.

Fitting of Raman data was performed with custom python code leveraging lmfit\cite{Newville2014} and scipy.\cite{Virtanen2020}
Expressions for the PO-dependence of Raman scattering from birefringent crystals in point group 2 were derived and are given in the supporting information, Section~10.

\textit{Impedance spectroscopy}: Preparations and measurements were performed in an argon glovebox. 5~mm diameter pellets with thicknesses between 1.5~mm and 2.5~mm were pressed uniaxially from powder and loaded into a TSC Battery measurement cell equipped with a Peltier element and temperature controller (RHD Instruments), where they were contacted under mild uniaxial pressure ($\sim$10~MPa) with stainless steel electrodes.
AC impedance measurements were performed (100~mV RMS excitation) with a Novocontrol NEISYS impedance analyzer.
For temperature studies, samples were equilibrated for an hour at each setpoint before measurement.
Sample impedance was modeled as the series combination of a parallel R$-$C circuit (representing the bulk) and a parallel R$-$Q circuit (representing grain boundaries; Q = constant phase element). For low-resistivity CsSnBr$_3$, the contact resistance (of order 10~$\Omega$) and parasitic inductance of the leads (of order 100~nH) are significant and were included in the model.

\textit{Solid state nuclear magnetic resonance spectroscopy}: $^{119}$Sn solid state NMR experiments were performed using a Bruker Avance-III 400 MHz spectrometer utilizing a Bruker BL 4 mm double-resonance magic-angle spinning (MAS) probe and a zirconia rotor, spun at 14~kHz. Spectra are referenced to (CH$_3$)$_4$Sn using SnO$_2$ as a secondary reference ($^{119}$Sn $\delta_{iso} = -603.0$~ppm). Due to the fast longitudinal relaxation, spectra could be acquired with a recycle delay of 10~ms. Single pulse spectra were collected for 2$^{14}$ or 2$^{15}$ scans, and spin-echo experiments for 2$^{10}$ scans. The spin-echo experiments employed a single rotor-period delay ($\sim$7.1~$\mu$s).

\section{Supporting Information}

Crystal structure models computed with a screened hybrid functional, additional experimental and computational data and analyses

\section{Acknowledgements}
D.H.F. thanks Hanna Bostr{\"o}m, Andreas Leonhardt, and Eric Riesel for helpful discussions, Frank Adams for technical assistance with cryogenic laboratory powder diffraction, and Igor Moudrakovski for assistance with design of the spin echo experiment.
D.H.F. thanks Prof. Rainer P{\"o}ttgen for provision of the M{\"o}ssbauer instrument and for helpful symmetry discussions.
D.H.F., S.B., and B.V.L. thank Robert Dinnebier for helpful discussions.
Use of the Advanced Photon Source at Argonne National Laboratory was supported by the U. S. Department of Energy, Office of Science, Office of Basic Energy Sciences, under Contract No. DE-AC02-06CH11357.
At Northwestern, work was supported by the U.S. Department of Energy, Office of Science, Basic Energy Sciences, under award number DE-SC-0012541 (synthesis and fundamental studies of metal halides).
D.H.F. gratefully acknowledges financial support from the Alexander von Humboldt Foundation and the Max Planck Society.

\providecommand{\latin}[1]{#1}
\makeatletter
\providecommand{\doi}
  {\begingroup\let\do\@makeother\dospecials
  \catcode`\{=1 \catcode`\}=2 \doi@aux}
\providecommand{\doi@aux}[1]{\endgroup\texttt{#1}}
\makeatother
\providecommand*\mcitethebibliography{\thebibliography}
\csname @ifundefined\endcsname{endmcitethebibliography}
  {\let\endmcitethebibliography\endthebibliography}{}

\end{document}